\documentclass[apj]{emulateapj}
\usepackage{epsfig}
\usepackage{epstopdf}
\usepackage{graphics}

\begin{document}

%=====TITLE AND ABSTRACT=======================================================

\title[]{Spatial-Dependent Diffusion of Cosmic Rays and the Ratio of $\overline p/p$, B/C}
\author{Yi-Qing Guo, Zhen Tian, Chao Jin}
\affil{Key Laboratory of Particle Astrophysics, Institute of High Energy Physics, Chinese Academy of Sciences,Beijing 100049,China}

\begin{abstract}

  Recent precise measurements of cosmic ray spectral revealed an anomalous hardening at $\sim$200 GV for nuclei from
  PAMELA, CREAM, ATIC, AMS02 experiments and at tens of GeV for primary electron
  derived from AMS02 experiment. Particularly, the latest observation
  of $\overline p/p$ ratio by AMS02 demonstrated
  a flat distribution, which further validated the spectrum anomalies of secondary particles. All those new phenomena
  indicated that the conventional propagation model of cosmic rays meet challenge.
  In this work, the spatial-dependent propagation coefficient
  $D(r,z,\rho)$ is employed by tracing the
  source distribution under the physical picture of two-halo model in DRAGON package. Under such scenario,
  the model calculation will result in a two-component spectral for primary nuclei
  and electron. Simultaneously, due to the smaller rigidity dependence of $D(r,z,\rho)$ in galactic disk, the ratio of secondary-to-primary
  will be inevitablly flatter than the calculation in the conventional propagation model.
  As a result, we can reproduce the spectral hardening of proton, electron  and the flat ratio of $\overline p/p$ and $B/C$ by only
  adopting the spatial-dependent propagation coefficient $D(r,z,\rho)$ in galactic disk.

\end{abstract}

\maketitle

\section{Introduction}

  Great progress in Cosmic Ray (CR) spectrum measurement has been made in recent years with new generation
  of space borne and ground based experiments in operation.
  The fine structure of
  spectral hardening for primary nuclei at 200 GV was observed by ATIC-2 \citep{2006astro.ph.12377P},
  CREAM \citep{2010ApJ...714L..89A} and PAMELA \citep{2011Sci...332...69A}. Just recently, AMS02 also confirmed
  the hardening, though the spectral line-shape has a bit discrepancy with each other \citep{2015arXiv150404276G}.
  Several kinds of explanations have been proposed to understand the origin of the spectral hardening, including: the contribution
  from nearby SNRs \citep{2012MNRAS.421.1209T}, the re-acceleration mechanism of old SNRs sources \citep{2010ApJ...725..184B},
  the combination effects from different group sources
  \citep{2006A&A...458....1Z,2011PhRvD..84d3002Y} and the spatial-dependent diffusion of
  CRs \citep{2012ApJ...752L..13T,2015arXiv150400227G,2015arXiv150406903J}.

  Along with the acceleration of nuclei, the primary electron can also be accelerated to high energy. This hints that the spectral hardening
  should happen in the primary electron. To reconcile the AMS02 positron fraction and total $e^\pm$ spectral
  observed by Fermi-LAT/HESS, the hardening spectral for primary electron was predicted by the work \citep{2013PhLB..727....1Y,2015APh....60....1Y}.
  Soon after the AMS02 publication of the $e^\pm$ spectral \citep{2014PhRvL.113l1102A}, the primary electron flux
  was derived from the substraction of positron
  flux as $\Delta \phi = \Phi_{e^-} - \Phi _{e^+}$ \citep{2014arXiv1412.1550L}.
  The most interesting thing was that the spectrum index shows roughly constant
  character \citep{2014arXiv1412.1550L} above tens of GeV, which was conflictive with the softening variation with energy
  in the Conventional Propagation
  Model (CPM) \citep{2014arXiv1412.1550L,2014arXiv1409.6248L}.
  After subtracting the flux from the calculation in CPM, the excess was uncased with the peak energy
  at $\sim$100 GeV \citep{2014arXiv1412.1550L}. In one words, more and more studies accepted the spectrum hardening for the primary electron
  with recent high precise measurement at high energy.
  The possible twinborn origin with nuclei has been proposed from the spatial-dependent
  propagation under the physical schema of Two-Halo-Model(THM) \citep{2012ApJ...752L..13T} in our previous work \citep{2015arXiv150406903J}.

  %Owing to the spectrum hardening at hundreds of GeV, the secondary particles, produced in the interaction between CRs and
  %InterStellar Medium (ISM), should exhibit anomalous property to the conventional propagation model.
  Similar phenomena also happened in the secondary particles.% which were possibly produced in the interaction between CRs and
  %InterStellar Medium (ISM).
  The ratio of $\overline p/p$ and B/C
  are tagged to be the characteristic quantity to calibrate the propagation of CRs. The predicted $\sim E^{-\delta}$ for secondary-to-primary
  lead to a sharp softening with energy for the ratio of $\overline p/p$ and B/C in the CPM.
  %and $E^{-\delta}$ for secondary-to-primary ratios, where $\beta$ is the injection spectrum index and $\delta$ is
  %the Kolmogorov spectrum in the interstellar turbulences with the value 1/3 \citep{1995ApJ...438..763G}.
  Considering the uncertainty, the previous
  result of $\overline p/p$ from PAMELA experiment seemed to be compatible with the CPM at the energy
  range of GeV to tens of GeV \citep{2014PhysicsReport,2010PhRvL.105l1101A}. Thanks to AMS02 experiment, the ratios of $\overline p/p$
  and B/C are measured from GeV energy to hundreds of GeV with high precision \citep{2015arXiv150404276G}.
  It is very exciting that the ratio of $\overline p/p$ is almost flat from $\sim$ 10 GeV to hundreds of GeV, which obviously
  challenge the CPM. This new result has stimulated several theoretical studies with point of view from either exotic
  physics \citep{2015arXiv150407230L,2015arXiv150504031C,2015PhRvD..91i5006G,2015arXiv150606929H,2015PhRvD..91k1701I,2015arXiv150700828C,2015PhLB..747..495C,2015PhLB..747..523H,2015arXiv150806844J} and some authors proposed that the slower diffusion in high energy can also result in the
  $\overline p$ excess \citep{2015arXiv150404276G,2015arXiv150500305C,2015arXiv150604145K}.

  With all of these high precision measurements in hand, one should ask whether a unified physical mechanism can result in those anomalous
  phenomena. A phenomenological model has been proposed to explain those anomalies by adopting the Hard Galactic Plane Component (HGPC)
  and predicted the flat distribution of $\overline p/p$, which was highly consistent
  with the later observation of AMS02 \citep{2014arXiv1412.8590G}.
  %However, the hadronic interactions are hand-input in the source region and the propagation of
  %primary harder spectrum component CRs is missing.
  One possible mechanism of the HGPC is the smaller rigidity dependence of diffusion in the galactic plane, such as THM.
  Inspired by this work,
  The optimal bedrock is from the understanding of CPM and compensate the missing parts in the CPM.
  The diffusive properties were assumed to be the same and this hypothesis leads to a uniform spectrum index
  for primary particles and their production of secondary ones in the whole
  galaxy in the CPM. A natural solution is to consider the spatial-independent propagation.
  One ready example is the THM, proposed by Tomassetti \citep{2012ApJ...752L..13T}, where the the propagation volume was divided into
  two regions as Inner Halo (IH) and Outer Halo (OH). The key point of THM is that the diffusion coefficient has a smaller
  rigidity dependence in the thin IH than in the wide OH, which can result in the a two-component spectrum.
  In 1 D analytical calculation, the spectrum hardening of nuclei spectra at rigidity of $\sim$200 GV,
  observed by ATIC-2 \citep{2006astro.ph.12377P},
  CREAM \citep{2010ApJ...714L..89A} and PAMELA \citep{2011Sci...332...69A}, were successfully reproduced \citep{2012ApJ...752L..13T}.
  Following this picture, Jin et al. extend it to 2D space and explain the spectrum hardening of primary electron \citep{2015arXiv150406903J}.
  %However, $\delta$ is also only Z-dependent, which result in a smooth excess for primary electron \citep{2015arXiv150406903J}.
  %Furthermore, the new observed $\overline p/p$ and $B/C$ ratio are not finished in that work.
  %The latest observation of $\overline p/p$ exhibited a flat distribution and conflict with the CPM.
  %It is necessary to continue further study to explain the new observation of $\overline p/p$ and $B/C$.
  The THM can work well in explanation of the spectral hardening of the primary CRs.
  It is necessary to continue further study to examine the secondary production of the new observation of $\overline p/p$ and $B/C$.

  The IH, actually called disk, contain almost all the galactic astrophysical object
  , which can generate large irregularities of the turbulence and play important impact to the diffusion properties.
  This imply that the diffusion coefficients is possibly related to the source distribution.
  On the contrary, the OH has scarce active sources and the dominant contribution of turbulence
  come from the CRs themselves \citep{2012ApJ...752L..13T}. This means that the diffusion coefficients can
  be fixed to constant. Such physical schema was consistent with the diffuse $\gamma$-ray observation
  by Fermi-LAT \citep{2012ApJ...750....3A}, which shows that the excess over the prediction of CPM is only existed in
  the galactic plane. Furthermore, the observed power-law index ranges from $E_\gamma^{-2.47}$ to $E_\gamma^{-2.6}$, resulting
  in a more harder spectrum in more close to the GC \citep{2015PhRvD..91h3012G}. The spatial diffusion coefficients should
  also depend on the radial coordinate.
  This paper tend to reproduce the spectrum hardening of primary particles and simultaneously reproduce the ratios
  of $\overline p/p$, $B/C$ by introducing the spatial-dependent diffusion parameters, which derived by tracing the source distribution.

  %Earlier,
  %to interpret PAMELA positron excess, the hadronic interaction and subsequent acceleration
  %for secondary particles around source region were proposed. Under such scenario, the excesses of $\overline p/p$ and $B/C$
  %ratio were expected
  %\citep{2009PhRvL.102e1101A,2009PhRvL.103h1103B,2009PhRvL.103h1104M,2009PhRvD..80l3017A,2014PhRvD..90f1301M}.
  %In addition, a self-consistent model has been proposed to link the excesses for most of the secondary particles by "fresh" CRs
  %\citep{2014arXiv1412.8590G}. However, the hadronic interactions are hand-input in the source region and the propagation of
  %primary fresh CRs is not considered. Following our previous work to interpret the spectrum hardening for primary nuclei and
  %electron by adopting spatial dependent diffusion, this paper tend to reproduce the ratios of $\overline p/p$ and $B/C$.
  %Different from our previous work, the spatial-dependent diffusion parameters are derived by tracing the source distribution in this paper.

  The paper is organized in the following way. Section 2 describes the spatial-dependent propagation of CRs,
  Section 3 presents the results of the calculation compared with the observation.
  Finally, Section 4 gives the conclusion.

\section{Spatial-Dependent Propagation of CRs}

  %The CRs before arriving at earth had gone through three stages on acceleration, propagation and solar moduation.
  During the active phase of astrophysical object
  such as SNRs \citep{1978MNRAS.182..147B,1978MNRAS.182..443B,1978ApJ...221L..29B},
  Galactic Center \citep{1981AZh....58..959P,1981ICRC....2..344S,1983JPhG....9.1139G,2013NJPh...15a3053G} and other ones,
  the expanding diffusive shocks are generated and can accelerate the
  CRs to very high energy.
  Before arriving at earth, those CRs have traveled in the galaxy for $\sim 10^7$ years
  after they diffuse away from the acceleration site \citep{1977ApJ...217..859G}.
  During the journey, the impacts due to the fragmentation and radioactive decay in the ISM result
  in the production of secondary particles. Meanwhile, the electron suffers energy loss in the interstellar radiation field (ISRF)
  and magnetic field.
  The story to experience in the long journey can be described by the propagation equation as:
\begin{equation}
\begin{array}{lcll}
  \frac{\partial \psi(\vec{r},p,t)}{\partial t} &=& q(\vec{r}, p,t) + \vec{\nabla} \cdot
                                     \left( D_{xx}\vec{\nabla}\psi - \vec{V_{c}}\psi \right)\\
                                   &+& \frac{\partial}{\partial p}p^2D_{pp}\frac{\partial}{\partial p}\frac{1}{p^2}\psi
                                   - \frac{\partial}{\partial p}\left[ \dot{p}\psi - \frac{p}{3}
                                     \left( \vec{\nabla}\cdot \vec{V_c}\psi \right) \right]\\
                                   &-& \frac{\psi}{\tau_f} - \frac{\psi}{\tau_r}
\end{array}
\label{CRsPropagation}
\end{equation}
  %Three stages of the evolution for GCRs are also sketched out in this formula base on above discussion.
  where $\psi(\vec{r},p,t)$ is the density of CR particles per unit momentum
  $p$ at position $\vec{r}$; $\vec{V_c}$ is the convection velocity;
  $\tau_f$ and $\tau _r$ are the characteristic time scales for fragmentation and
  radioactive decay respectively; $q(\vec{r}, p,t)$ is the source distribution; $D_{xx}$ and $D_{pp}$ are the
  diffusion coefficients in coordinate and momentum space respectively.
  In this formula, three stages of the evolution for observed CRs are also sketched out as: injection, propagation
  and solar modulation.

  {\bf Injection Spectral: }
  Though many kinds of astrophysical objects can accelerate the CRs to very high energy, SNRs have long been considered as
  the dominant candidates of GCRs. The accelerated spectrum of primary CRs at source region is assumed to
  be a broken power law function and the formula is similar as precious works \citep{2014arXiv1412.8590G,2015arXiv150406903J}.
  At high energy, the exponent cut-off at $\sim 4$ PeV is adopted for nuclei and a broken power law at $\sim$TV
  with power law index -4.1 is selected to agree with HESS and VERITAS observation \citep{2008PhRvL.101z1104A,2015arXiv150806597S}.
  Detailed information of the parameters is listed in Table \ref{PrimaryCRs}.
  \begin{table}[h]
  \begin{center}
  \caption{ The inject spectrum of primary CRs}
  \begin{tabular}{ccccccccccc}\hline \hline
   parameters     &&&&& Nuclei  &&&&& Electron  \\
   log($q^i_0$)   &&&&& -8.31 &&&&& -9.367     \\
   $\nu_1$        &&&&&  1.9    &&&&& 1.86       \\
   $\nu_2$        &&&&&  2.39   &&&&& 2.76       \\
   $p_{br}(GV)$   &&&&&  9.5    &&&&& 4.2       \\ \hline
   \end{tabular}
   \label{PrimaryCRs}
   \end{center}
   \end{table}

  {\bf Propagation Coefficients:}
  The spatial diffusion coefficient was a power law form as
  $D_{xx}$ = $D_0\beta (\rho/\rho_0)^\delta$, where $\rho$ is the rigidity and
  $\delta$ reflects the property of the ISM turbulence.
   The re-acceleration can be described by the diffusion in momentum space
  and the momentum diffusion coefficient $D_{pp}$ is coupled with the spatial
  diffusion coefficient $D_{xx}$ as \citep{1994ApJ...431..705S}
 \begin{equation}
   D_{pp}D_{xx} = \frac{4p^2v_A^2}{3\delta(4-\delta ^2)(4-\delta)w}
\end{equation}
  here $v_A$ is the Alfven speed, $w$ is the ratio of magnetohydrodynamic
  wave energy density to the magnetic field energy density, which can be
  fixed to 1. The CRs propagate in an extended halo with a characteristic
  height $z_h$, beyond which free escape of CRs is assumed.

  Based on the discussion about the THM in introduction, the spatial-dependent diffusion coefficients can be described as:

  \begin{equation}
    D(r,z,\rho) = \left \{
        \begin{array}{lll}
            D_0^\prime(r,z)\beta (\frac{\rho}{\rho_0})^{\delta(r,z,\rho)} &|z|<\xi z_h & (IH)\\
            D_0\beta (\frac{\rho}{\rho_0})^{\delta_0}                  &|z|>\xi z_h & (OH)
        \end{array}
           \right.
  \end{equation}
  where $\xi$ is the fraction of the half thickness of IH to the halo height $z_h$.
  The diffusion coefficients is dependent on the irregular turbulence, which possibily arises from
  the violent activity of astrophysical object,
  such as SNRs, pulsars, stellar winds, OB stars and so on \citep{2013APh....42...70E}. That is to say that
  the diffusion coefficients can be derived by the source distribution.

  %It is noticed that the break positron and altitude are dependent on $\xi$ and the difference of $\Delta \delta$ = $\delta_0$ - $\delta$.
  %According to above discussion, the diffusion coefficient should
  %be radial and Z dependent in the inner halo. In addition, It should be related to the souce distribution.

  %The violent activity of astrophysical object, such as SNRs, pulsars, OB stars and so on, may generate large irregularities
  %of the turbulence and magnetic-field. The random scattering on irregular turbulence region caused the diffusion of CRs.
  %The diffusion coefficients should depend on the sources. Fig. \ref{fig:Source} show the radial distribution of astrophysical object,
  %such as pulsar, SNRs and NS \citep{2004A&A...422..545Y}. Though the radial distribution become slow down toward GC,
  %the irregularity of turbulence and magnetic-field is more larger in more near the GC region. So the red line is the extended distribution
  %to GC region, which can be called the scale-line of the irregularity and served as the scalar of diffusion coefficients. In the halo,
  %the galactic source is very sparse and can be neglectful. The dominant contribution of turbulence come from the CRs themselves
  %\citep{2012ApJ...752L..13T} and the diffusion coefficients can be dealed with constant resonablly.

\begin{figure}[!htb]
\centering
\includegraphics[width=0.48\textwidth]{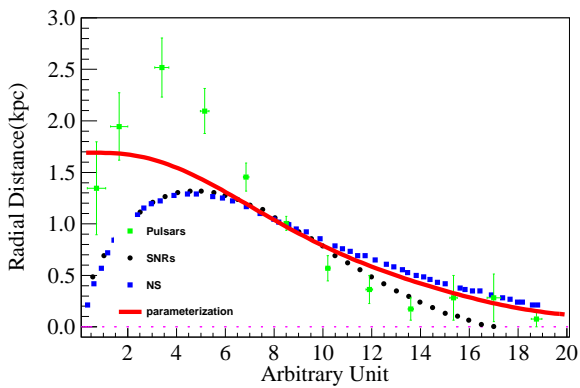}
\caption{Radial distribution of astrophysical objects: Pulsar, SNRs and NS \citep{2004A&A...422..545Y}. The red line is the
         "scale-line", derived from the combined source distribution and extended to the GC.}
\label{fig:Source}
\end{figure}

\begin{figure*}[!htb]
\centering
\includegraphics[width=0.47\textwidth,height=0.42\textwidth]{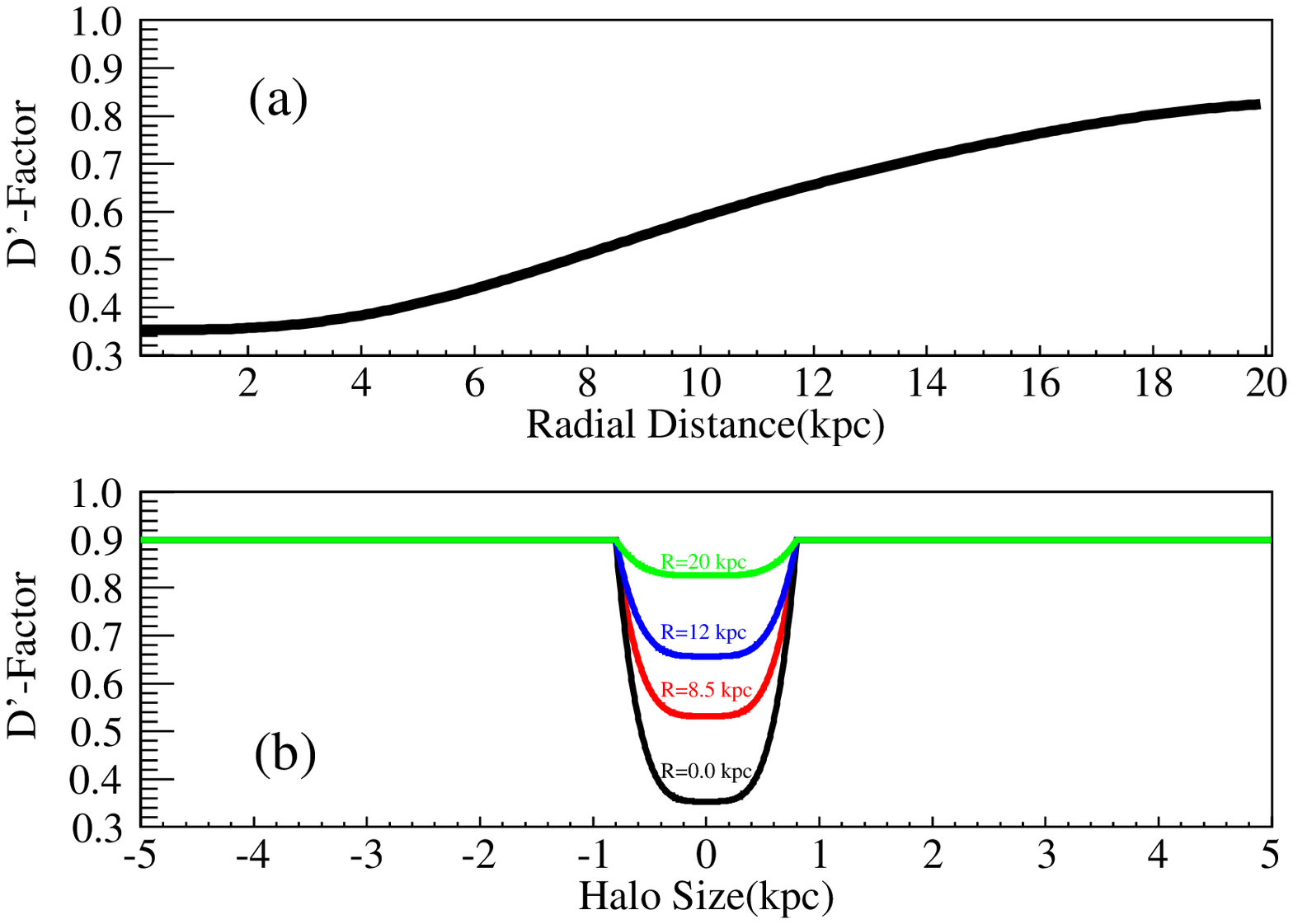}
\includegraphics[width=0.47\textwidth,height=0.42\textwidth]{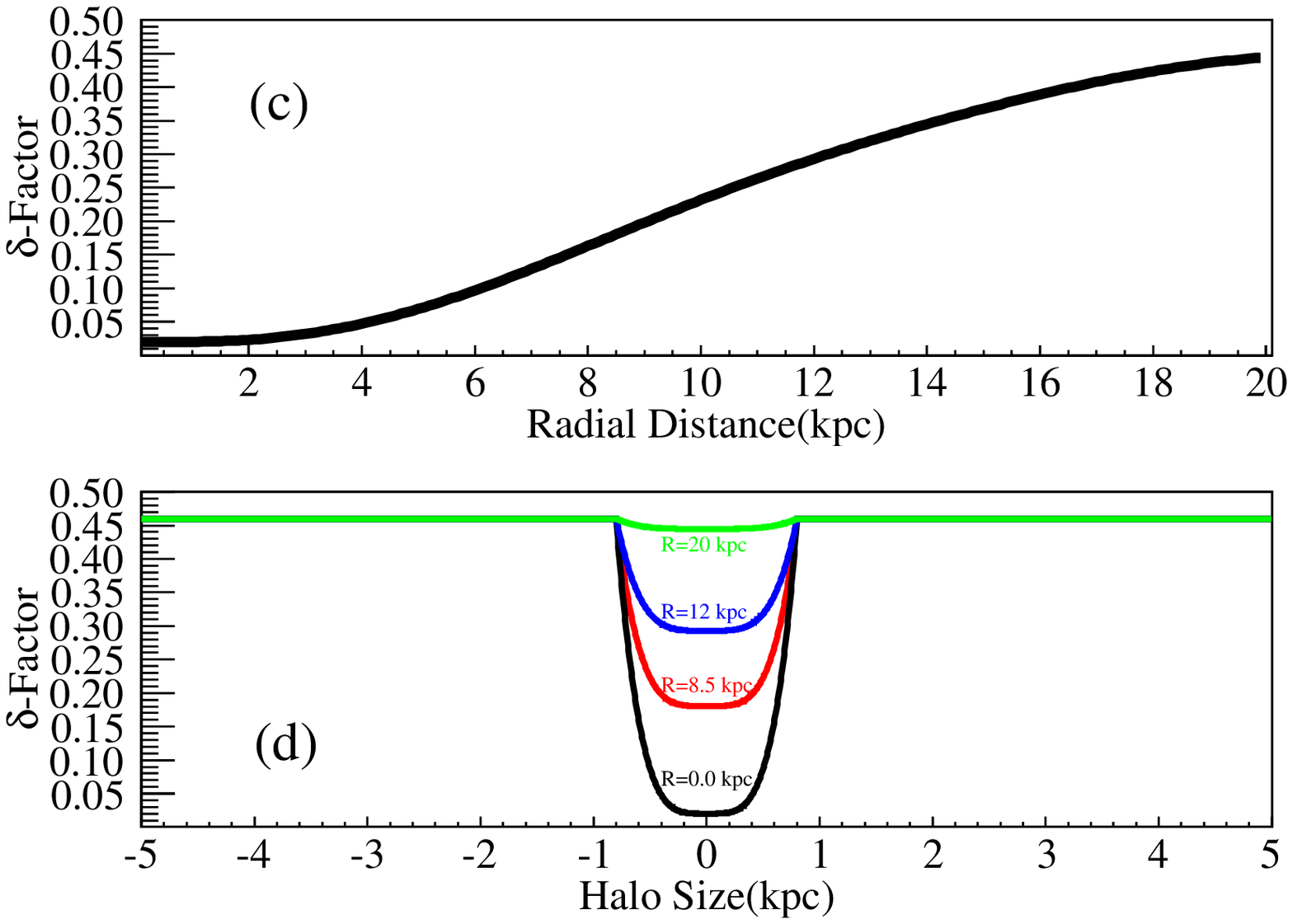}
\caption{The distribution of spatial-dependent diffusion coefficency: D(r,z) and $\delta$(r,z) in radial and z direction,
         here (a) is the D$^\prime$-Factor along radial coordinate; (b) is D$^\prime$-Factor along Z coordinate at
	 different radius; (c) is $\delta$-Factor along
         radial coordinate and (d) is $\delta$-Factor along Z coordinate at different radius.}
\label{fig:PropProton}
\end{figure*}

 Fig. \ref{fig:Source} shows the radial distribution of astrophysical object: pulsar, SNRs and NS \citep{2004A&A...422..545Y}.
 Though the radial distribution become slow down toward GC,
 the irregularity of turbulence and magnetic-field is more larger in more near the GC region.
 Evidences that there is an increase in turbulence as one approaches the GC are provided by observations of
 magnetic field \citep{2009ASTRA...5...43B} and thermal bremsstrahlung \citep{2011A&A...536A..21P}.
 Combined with the source distribution and the magnetic field information toward GC, the irregularity turbulence along radius can
 be built as the red solid line, here we call it "scale-line".
 It is true that the shape of scale-line is a bit subjective. However, the scale-line is only served to describe
 the radial dependent diffusion property, which can be determined by the spectral observations as following discussion.
 %The red solid line is the extended distribution to GC region,
 %which can be called the scale-line of the irregularity and served as the scalar of diffusion coefficients in the IH.
 Furthermore,
 the galactic source is very sparse and can be neglectful in OH.
 The dominant contribution of turbulence come from the CRs themselves
 \citep{2012ApJ...752L..13T} and the diffusion coefficients can be dealed with constant reasonably.
 The diffusion coefficients of $D_0^\prime(r,z)$ and $\delta(r,z)$ can be described in a unified
 formula as:
\begin{equation}
  F(r,z) = \left \{
        \begin{array}{ll}
           (A/(1+exp(f(r))) - B)(z/\xi z_t)^n & (IH)\\
           C                        & (OH)
        \end{array}
           \right.
  \label{Dconstruction}
\end{equation}
 where f(r) is the scale-line; the factor $(z/\xi z_t)^n$ is used to describe the smooth connection 
 between two halos and can be fixed to 4 (similar with the work \citep{2015arXiv150406903J});
 A, B, C are free parameters, which can be fixed by
 fitting the observation spectral.
 %In this work, the values are listed in Table \ref{TabProp}:
 %Formula \ref{Dconstruction} seems very complicated, but the
 %intention is very simple and just to get the quantitative and radial-dependent diffusion coefficient together with the scale-line.
 %Although the expression of Formula \ref{Dconstruction} is little complicated than
 %Formula \ref{Dconstruction} only intend to give a quantitative and radial-dependent diffusion coefficient together with the scale-line.
 Formula \ref{Dconstruction} seems a little complicated.
 Actually, we can also select the simplest format $\delta(r) = A\bullet r + B$, but
 need to deal with the region close and far (r$>$11 kpc) from GC to avoid the saturation as the work \citep{2015PhRvD..91h3012G}.
 Here we just intend to get a quantitative and smooth diffusion coefficient along the radial coordinate and
 adopt the formula \ref{Dconstruction}. By adopting the values of the parameters as showing in Table \ref{TabProp},
 Fig. \ref{fig:PropProton} shows the radial and Z-direction distribution for $D^\prime$ and $\delta$ respectively.
 %The parameters can be fixed by fitting the observation spectral and the values are listed in Table \ref{TabProp} in this work.
 %In this work, the parameters is adopted as Table \ref{TabProp}. Fig. \ref{fig:PropProton} shows the radial and Z-direction
 %distribution for D and $\delta$ respectively.
\begin{table}[h]
\begin{center}
\caption{The parameters of spatial-dependent diffusion.}
\begin{tabular}{ccccccccccc}\hline \hline
 parameters   &&&&& D$^\prime$(r,z) &&&&& $\delta$(r,z)  \\ \hline
 A      &&&&& 1.5          &&&&& 1.35     \\
 B      &&&&&  0.12        &&&&& -0.19       \\
 C      &&&&&  0.9         &&&&& 0.46       \\
 %n      &&&&&   4          &&&&& 4       \\
 $\xi$  &&&&&   0.16       &&&&& $\sim$ \\\hline \hline
\end{tabular}
\label{TabProp}
\end{center}
\end{table}

%The $D^\prime$(r) and $\delta$(r) are shown in
%Fig. \ref{fig:PropProton} respectively. % shows the radial and Z-direction distribution for $D^\prime$ and $\delta$ respectively.

{\bf Solar Modulation:} When the low energy CRs arrive at the solar environment, the direction of motion will be affected by the solar wind,
     which are called solar modulation. The force-field approximation is used to describe the solar modulation
     by the modulation potential $\Phi$ \citep{1968ApJ...154.1011G}. In this work, The modulation potential $\Phi$ is fixed as
     $\Phi$=500 MV for proton, $\overline p$, lepton and
     $\Phi$=200 MV for B/C.

\section{Calculation Results}

  In this calculation, we employ the publicly available numerical code DRAGON, which
  was designed to solve the spatial-dependent propagation of CRs \citep{2008JCAP...10..018E}.
  By adopting one set of diffusion coefficient as discussed in above, we perform the calculation of primary spectral for proton,
  electron and the ratio of secondary-to-primary for $\overline p/p$, B/C.

  %Based on the spatial-dependent diffusion coefficiency D(r,z) and $\delta$(r,z), the predicted spectral in observed reference system
  %for primary CRs can be approximately derived as:
  %\begin{equation}
  %   N_{sun} \sim \frac{L}{D}\left[ \xi p ^{-\nu-\delta} + (1-\xi)p^{-\nu-\delta_0}\right ]
  %   %\label{preSpectrum}
  %\end{equation}
  %where L= 5kpc is the half halo size; $\xi$ is the fraction of inner halo thickness to the total halo size and (1-$\xi$) is
  %corresponding to the fraction of outer halo size to the total halo size; $\delta$ is explicitly less than $\delta_0$
  %as shown in Fig. \ref{fig:PropProton}.
  %Two spectrum components are obviously extracted as: the sharp spectrum denoted by the term $(1-\xi)p^{-\nu-\delta_0}$
  %and the harder one denoted by the term $\xi p ^{-\nu-\delta}$.
  %The sharp spectrum will be dominant in the low energy with the fraction of (1-$\xi$), which can be imaged as the CRs in halo
  %backtrack to the disk. The harder spectrum will become important at high energy with the faction of $\xi$, which can be
  %understood to the fresh CRs around or extened the accelerators \citep{2014arXiv1412.8590G}.
  %In addition, the transition break is determ by the factor $\xi$. Similiar to the primary spectral, the spectral for
  %secondary particles should have inherited hardening properity. The predicted ratio of  $\overline p/p, B/C$ will become flatter
  %than the calculation results from the conventional propagation model.
  %Detailed comparisons between model calcualtion and  the observed results are described in the following subsections.

\subsection{The Spectral of primary proton and electron}

\begin{figure*}[!htb]
\centering
\includegraphics[width=0.47\textwidth]{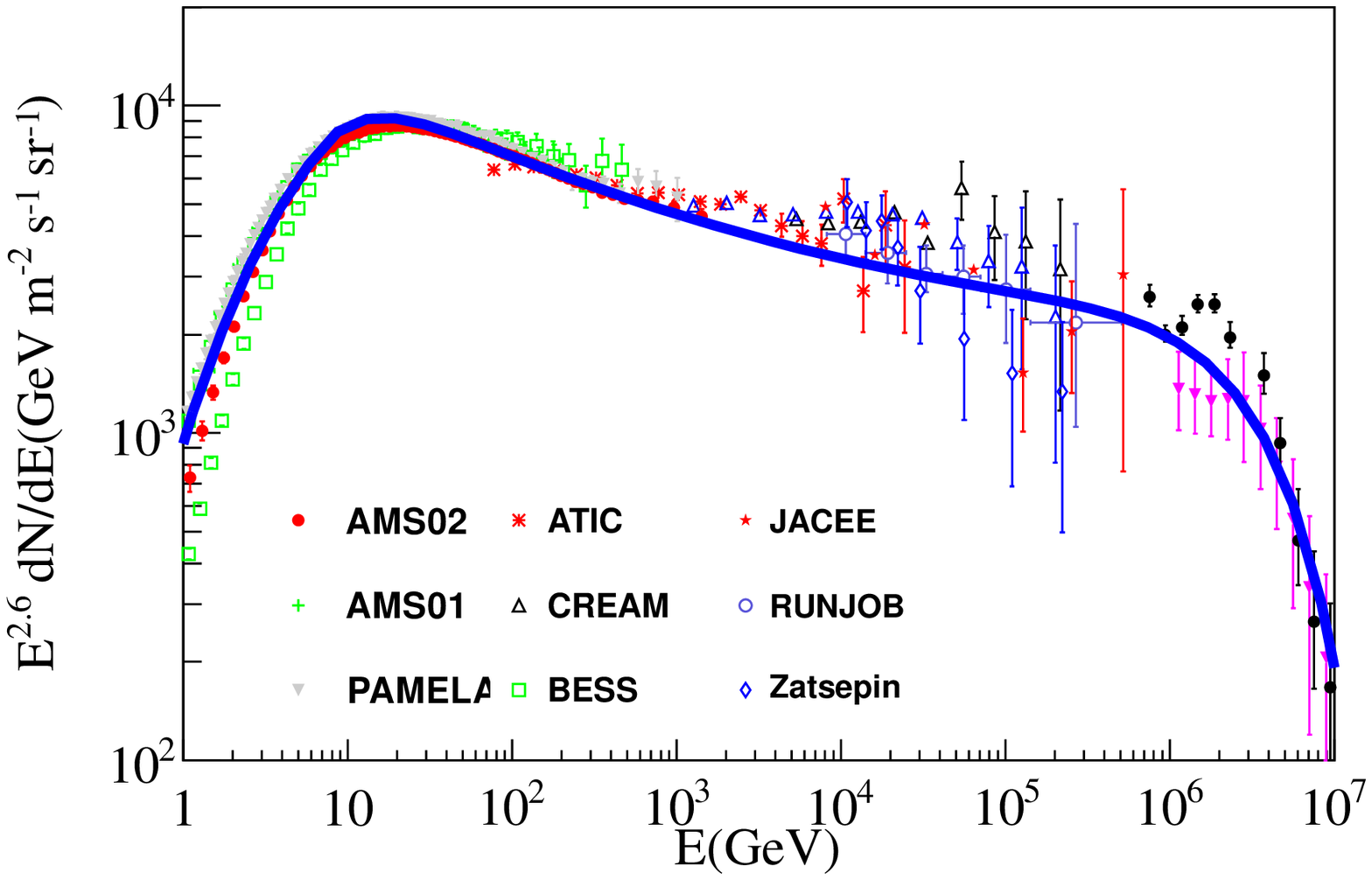}
\includegraphics[width=0.47\textwidth]{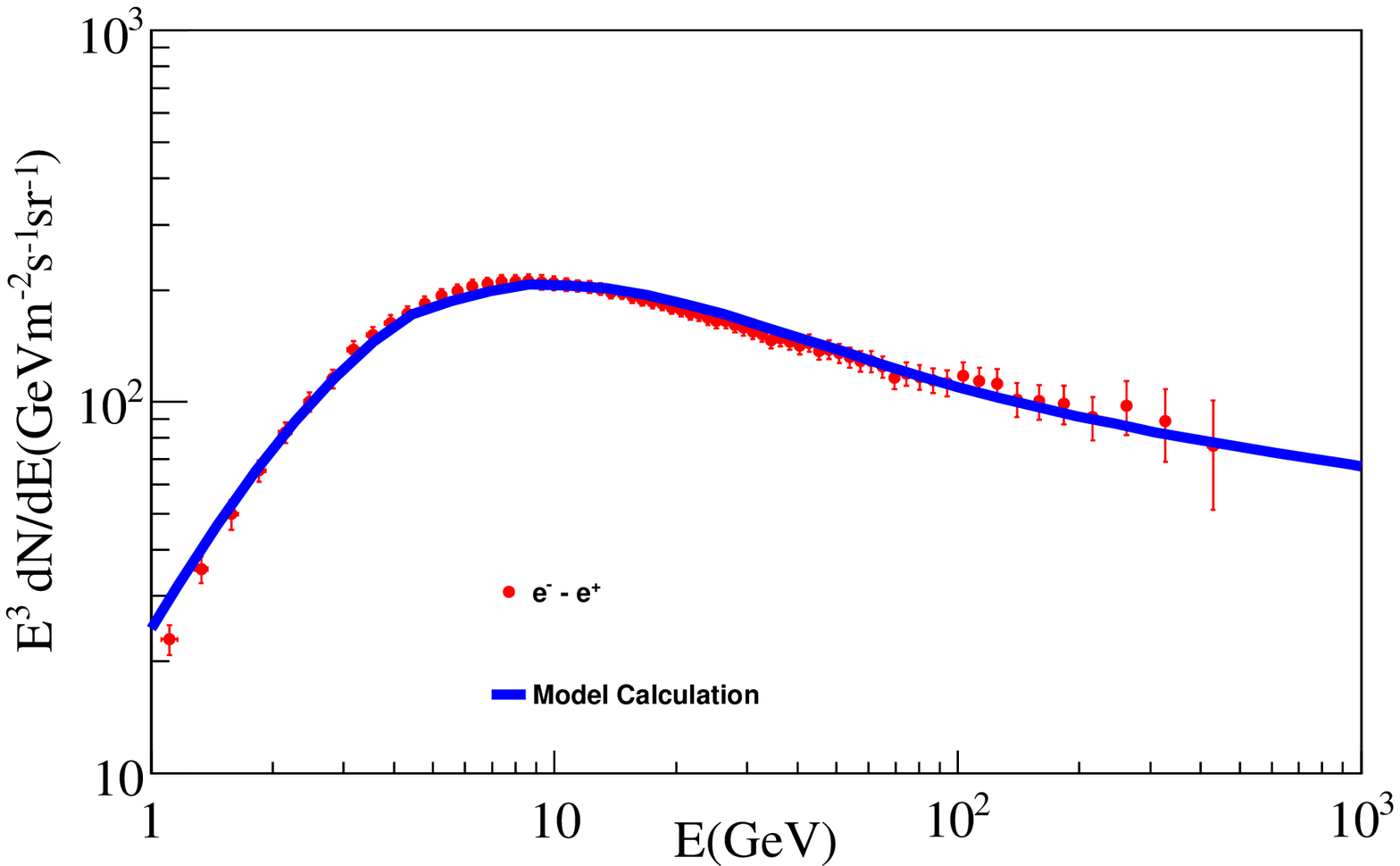}
\caption{The comparison between model calculations and observations for prmary spectral of proton and electron.
         The experiment data of proton come from: AMS02\citep{2015arXiv150404276G},
         ATIC\citep{2006astro.ph.12377P}, PAMELA\citep{2011Sci...332...69A}, AMS01\citep{2000PhLB..472..215A},
         CREAM\citep{2010ApJ...714L..89A},
         BESS\citep{2000ApJ...545.1135S}, JACEE\citep{1998ApJ...502..278A},
         RUNJOB\citep{2001APh....16...13A}, KASCADE\citep{2005APh....24....1A}. The $e^+e^-$ data come from
         \citep{2014PhRvL.113l1102A}.}
\label{fig:Primary}
\end{figure*}

The left panel of Fig.\ref{fig:Primary} shows the spectrum comparison between model calculation and
the experimental results for proton. The solid blue line represents the model calculation.
The spectrum break is clearly reproduced starting from hundreds of GeV, which are consistent with
the observation from AMS02 experiment, but a little lower than the observations of PAMELA and ATIC at TeV energy range.
According to above discussion, the break position and amplitude of spectrum is dependent on the
fraction $\xi$  and $\delta$ in the inner halo. High precise measurement in TeV energy range is expected to
determine those parameters in future.

  The left panel of Fig.\ref{fig:Primary} shows the spectrum comparison for primary electron.
  The flux of primary electron was derived from the subtraction of positron
  flux as $\Delta \phi = \Phi_{e^-} - \Phi _{e^+}$ \citep{2014arXiv1412.1550L} base on the AMS02
  observation of $e^\pm$ \citep{2014PhRvL.113l1102A}.
  It is obvious that the spectrum break is reproduced starting from tens of GeV and
  well consistent with the data.

  \subsection{The ratio of $\overline p/p$ and $B/C$}

\begin{figure*}[!htb]
\centering
\includegraphics[width=0.47\textwidth]{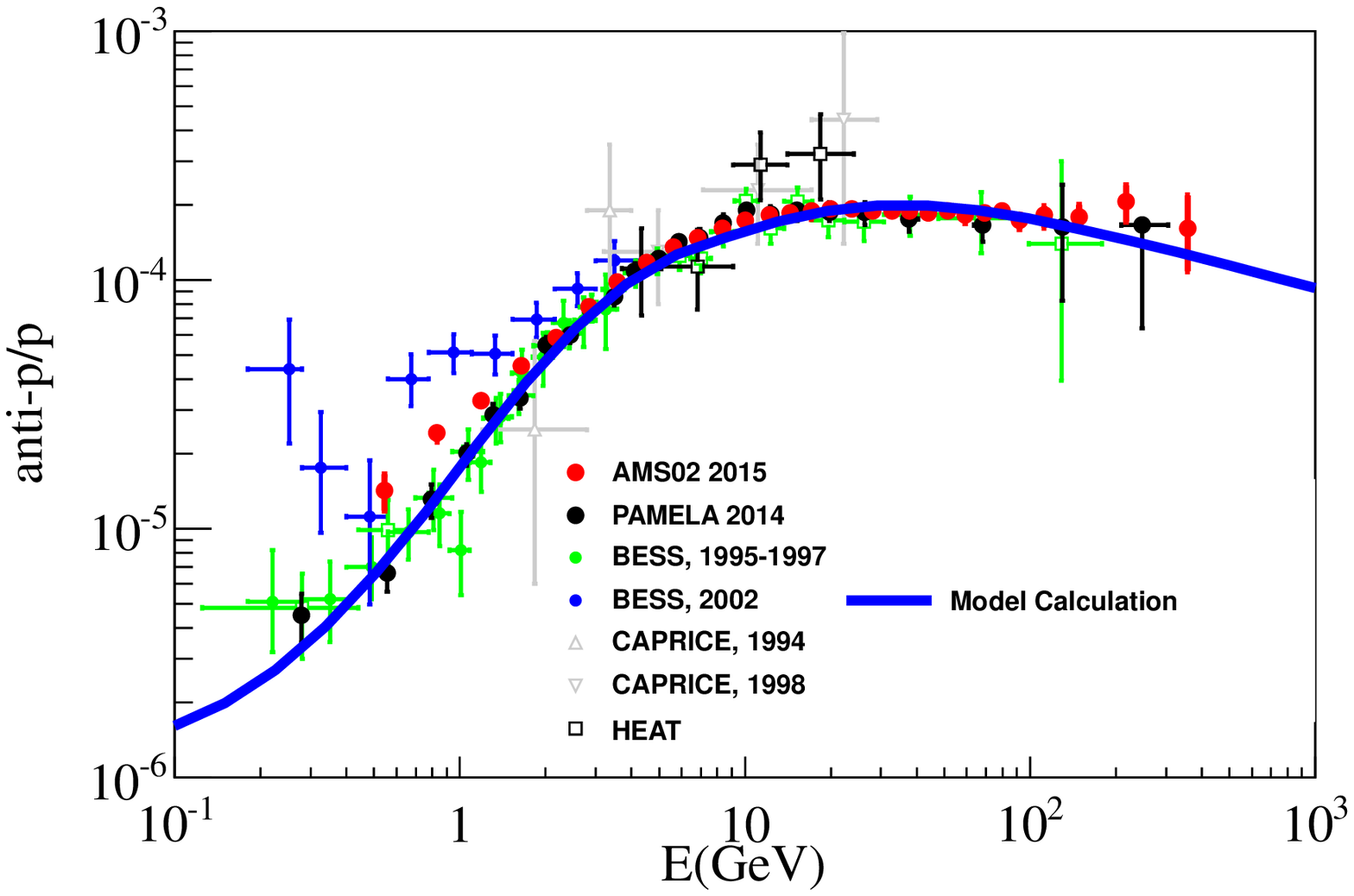}
\includegraphics[width=0.47\textwidth]{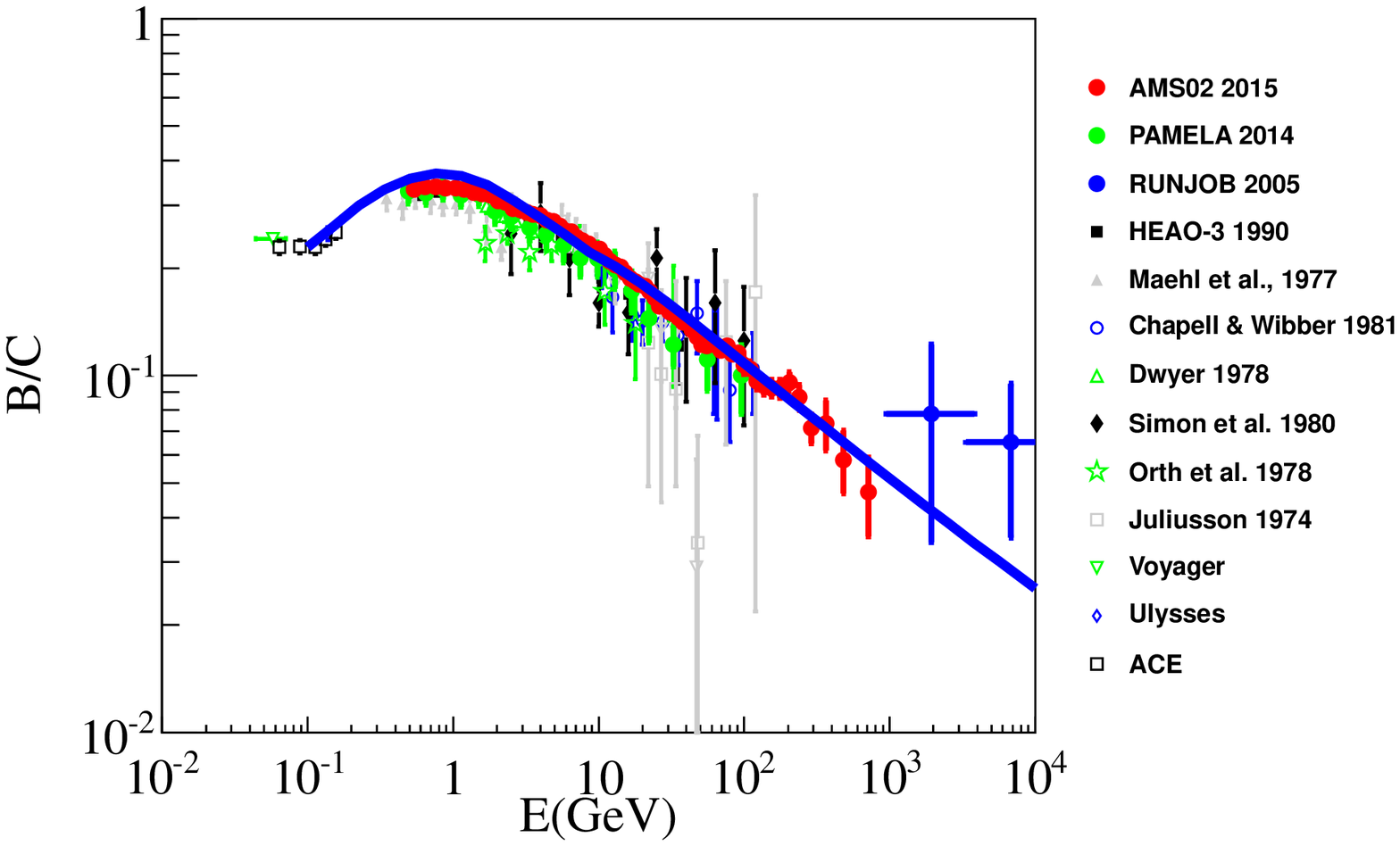}
\caption{The calculated $\overline p/p$ (left panel) and B/C (right panel). The $\overline p/p$ data
         from: AMS02 2015 \citep{2015arXiv150404276G}, PAMELA 2014 \citep{2014PhysicsReport},
         PAMELA 2010 \citep{2010PhRvL.105l1101A}, BESS 1995-1997 \citep{2000PhRvL..84.1078O}, BESS 1999 \citep{2002PhRvL..88e1101A},
         CAPRICE 1994 \citep{1997ApJ...487..415B}, CAPRICE 1998 \citep{2001ApJ...561..787B}, HEAT \citep{2001PhRvL..87A1101B}.
         The B/C data from: AMS02 \citep{AMS02BC}, PAMELA \citep{2014ApJ...791...93A}, RUNJOB \citep{2005ApJ...628L..41D},
         Juliusson \citep{1974ApJ...191..331J}, Dwyer \citep{1978ApJ...224..691D},
         Orth \citep{1978ApJ...226.1147O}, Simon \citep{1980ApJ...239..911S},
         HEAO-3 \citep{1990A&A...233...96E}, Maehl \citep{1977Ap&SS..47..163M}, Voyager \citep{1999ICRC....3...41L},
         Ulysses \citep{1996A&A...316..555D}, ACE \citep{2000AIPC..528..421D} and for other references see \citep{1998ApJ...505..266S}.}
\label{fig:BC}
\end{figure*}

  The left panel of Fig.\ref{fig:BC} shows the calculated $\overline p/p$ and the right panel shows the $B/C$.
  The blue lines present the calculated results from the spatial-dependent propagation.
  The ratio of $\overline p$ is raised up at high energy and consistent with the measurement of AMS02 within errors.
  %in high energy and with
  %PAMELA at $\sim$GeV energy.
  Below $\sim$2 GeV energy, the model calculation is consistent with PAMELA and a little lower than the measurement of AMS02.
  %On the other hand, the ratio of $B/C$ is quite consistent with AMS02 observation at high energy and a little higher than AMS02
  However, the uncertainty of $\overline p$
  production cross section is $\sim 25\%$ in the energy
  range 0.1 - 100 GeV \citep{2001ApJ...563..172D,2009PhRvL.102g1301D}, which should lead to
  a same level of uncertainty in the ratio calculation.
  Taking all of those factors into account, the model calculation is consistent
  with the observation within the errors. %The TeV
  On the other hand,
  the ratio of $B/C$ is quite consistent with AMS02 observation, except a little higher at $\sim$ 1 GeV.

  Anyway,
  Owing to its hard spectrum component at inner halo, the CRs induced secondary $\overline p$ and Boron inherit a similar
  hard spectrum, which make the ratio of $\overline p/p, B/C$ considerably flatter than
  that from conventional model. Such a tendency is favoured in just recent $\overline p/p$ observation by AMS02 below TeV energy region.
  In the future, high statistic and TeV energy observation can offer
  a crucial and definitive identification of this model.

\section{Conclusion and Future Work}

  In this work, the spatial-dependent propagation parameter are dirived from the source distribution under the THM scenario.
  By adopting one set of parameters, the primary spectral hardening of proton and electron
  are reproduced following previous work \citep{2012ApJ...752L..13T,2015arXiv150406903J}.
  Simultaneously, the new observed $\overline p/p$ and $B/C$ ratios can be reproduced under this model.
  High statistic and TeV energy observation can offer a crucial and definitive identification of this model in future.

  Currently, we only carry out the calculation of charged CRs and comparison with observations in solar system.
  The $\gamma$-ray emission can be used as a direct probe of the CR indensities and spectra in distant location
  which can serve to determine the spatial-dependent diffusion coefficients. Actually,
  our parameter of $\delta(r)$ reflects this property as discussed in the work \citep{2013APh....42...70E,2015PhRvD..91h3012G}.
  Further studies about the $\gamma$-ray emission can be performed in our future work.

\section*{Acknowledgments}
  This work is supported by the Ministry of Science and Technology of
  China, Natural Sciences Foundation of China (11135010).

\emph{Note added}: While this paper was in preparation, one similar
papers on interpretation for the spectral hardening under the two-halo model scenario
appeared \citep{2015arXiv150905775T}.

\bibliographystyle{apj}
\bibliography{space}

\end{document}